\documentclass[12pt]{amsart}
\usepackage{amssymb}
\usepackage{amsfonts}
\usepackage{graphicx}
\usepackage{epstopdf}
\usepackage{hyperref}

\newcommand{\beq}{\begin{eqnarray}}
\newcommand{\eeq}{\end{eqnarray}}
\def\G {\Gamma}

\def\a{\alpha}

\def\gg {{\mathfrak g}}
\def\G {\Gamma}

\def\C {{\mathbb C}}
\def\Z {{\mathbb Z}}

\theoremstyle{definition}

\begin{document}

\begin{center}
{\Large \bf Diophantine equations, Platonic solids, \\ McKay correspondence, equivelar maps \\ and \\ Vogel's universality} \\
\vspace*{1 cm}

{\large H.M.Khudaverdian\footnote{khudian@manchester.ac.uk}
}

\vspace*{0.2 cm}

{\small\it School of Mathematics,, University of Manchester,
         Oxford road, Manchester M13 9PL, United Kingdom}

\vspace*{0.5 cm}

{\large  R.L.Mkrtchyan\footnote{mrl55@list.ru}
}
\vspace*{0.2 cm}

{\small\it Yerevan Physics Institute, 2 Alikhanian Br. Str., 0036 Yerevan, Armenia}

\end{center}\vspace{2cm}

{\scriptsize  {\bf Abstract.} We notice that one of the 
Diophantine equations, $knm=2kn+2km+2nm$, arising in the universality 
originated Diophantine classification of simple Lie algebras, 
has interesting interpretations for two different sets of signs of  variables.
 In both cases it describes "regular polyhedrons"  with $k$ edges in each 
vertex,  $n$ edges of each face, with total number of edges $|m|$, 
and Euler characteristics $\chi=\pm 2$.  
In the case of negative $m$ 
this equation corresponds to $\chi=2$ and 
describes true regular  polyhedrons, Platonic solids. 
The  case with positive $m$ corresponds to Euler characteristic 
$\chi=-2$ and describes the so called equivelar maps (charts)
on the surface of genus $2$. In the former case  there are two routes from 
Platonic solids to simple Lie algebras --- abovementioned Diophantine 
classification and McKay correspondence. We compare them for 
all solutions of this type, and find coincidence in the case of 
icosahedron (dodecahedron), corresponding to $E_8$ algebra.  
In the case of positive $k$, $n$ and $m$ 
we obtain in this way the  interpretation of (some of) the 
mysterious solutions (Y-objects), appearing in the 
Diophantine classification and having some similarities with simple Lie algebras.}

\section{Introduction}

Since ancient time  natural numbers have been  suggested 
as a basic notion in the construction of our knowledge about nature. 
However, it is rare when they are the part of the
 basis of construction of a given mathematical or physical theory. In this paper we consider several such cases and observe that they are based on the same Diophantine equation. Moreover, since two of that cases are connected to simple Lie algebras, we are naturally led to comparison of that two superficially disjoint theories. 

The focus of the present paper is  the following Diophantine equation
\begin{eqnarray} \label{Dio}
\frac{1}{k}+\frac{1}{n}+\frac{1}{m}=\frac{1}{2}\,, \\ \nonumber
k,n,m \in \mathbb{Z} \setminus 0
\end{eqnarray}
or in more general form, which allows zero values of integers $k, n, m$:

\begin{eqnarray} \label{Dio2}
knm=2kn+2km+2nm\,.
 \end{eqnarray}
We would like to point out that this equation
 appears in three circumstances, 
depending, particularly, on the signs of integers $k,n,m$.
In two of them there are (different)  routes from this equation to 
simple Lie algebras. 
 
   One route is the famous McKay correspondence \cite{McK80}.
 It is well-known 
that solutions of equation (\ref{Dio}) with $(++-)$ signs of variables 
describe  Platonic solids (see below in Section \ref{PlaS}).
 Take invariance subgroup of given Platonic solid 
(it is finite subgroup of the group $SO(3)$, lift it to group $SU(2)$ 
by double-covering map
\begin{eqnarray}
1 \rightarrow \Z_2\rightarrow SU(2) \rightarrow SO(3) \rightarrow 1\,,
\end{eqnarray} 
and assign to this subgroup of $SU(2)$
by McKay procedure the simple Lie algebra from the list of ADE 
algebras (see Section \ref{sectM}). 
Note that one has to consider also degenerate "Platonic solids", 
and take into account different liftings of groups. All that is 
briefly described in Section \ref{sectM}.

Other route from Diophantine equation (\ref{Dio}), 
with the same set of signs, to simple Lie algebras  is given by
recently developed \cite{M16} Diophantine classification of 
simple Lie algebras, based on Vogel's universality \cite{V0,V2} and Deligne's conjecture on exceptional simple Lie algebras \cite{Del}. This is  briefly described in Section \ref{sectD}.

In section \ref{sectComp}    we  compare these 
two routes from solutions of  Diophantine equation
(\ref{Dio}) to simple Lie algebras, 
and find several common features and differences.

Finally, we discuss relation of Diophantine equations
(\ref{Dio}) with the 
theory of equivelar maps \cite{McS02,E22,T32} on orientable surfaces of genus two.
They appear to correspond to the same equation (\ref{Dio}) 
with $(+++)$ signs of variables. 
In Diophantine classification this case corresponds to 
mysterious $Y$-objects, which have certain similarity with 
simple Lie algebras, but up to now were not 
identified with any known objects. This is discussed in 
Section \ref{sectRM}.

\section{Platonic solids' Diophantine equation} \label{PlaS}

Consider Platonic solid with number of edges of any face  
$r$, number of edges at any 
vertex  $n$,  total number of edges  $E$, 
total number of vertices  $V$, and total number of faces  $F$. 
We have
\begin{equation*}\label{vef1}
nV=2E\,,\quad  rF=2E\,.
\end{equation*}
Then Euler's theorem 
\begin{equation*}\label{eulertheorem1}
V-E+F=2
\end{equation*}
can be rewritten as 
\begin{equation} \label{pla1}
\frac{1}{r}+\frac{1}{n}-\frac{1}{E}=\frac{1}{2}\,.
\end{equation}
This  is the particular case of Diophantine equation (\ref{Dio}) 
with the special choice $(++-)$ of signs of integers $k, n, m$. 

Solutions $(r,n,E)$ of equation (\ref{pla1})  are:

\begin{itemize}

\item
$(5,3,30)$ or $(3,5,30)$ --- dodecahedron or icosahedron,

\item $(4,3,12)$ or $(3,4,12)$ --- cube (hexahedron) or octahedron,

\item $(3,3,6)$ --- tetrahedron,

\item $(2,n,n)$ --- regular $n$-polygon.

\end{itemize}
This information is listed in the first and second column of 
table \ref{tab:s}.

\begin{scriptsize}
\begin{table}[ht] 
\caption{McKay correspondence and Diophantine classification.}
	\centering
		\begin{tabular}{|r|r|r|r|r|}  \hline
			\hbox {Solutions}&{Platonic }&
               Subgroups &
                  McKay & Diophantine\\ 
        (k,n,m)     & solids &    of $SU(2)$ &
                  correspondence &  classification \\ \hline
    (5,3,-30)   & Icosahedron  &  & & \\ 
     (3,5,-30)  &  Dodecahedron&$2I, |2I|=120$ &$E_8$ & $E_8$\\ \hline
	(4,3,-12) & Cube  & & & \\
	(3,4,-12)& Octahedron&$2O, |2O|=48$ &$E_7$ & $E_6$\\ \hline
   (3,3,-6)  & Tetrahedron&$2T, |2T|=24$ &$E_6$ & $SO(8)$ \\ \hline
	(2,n,-n)& $n$-polygon&$C_n, |C_n|=n$ &$A_{n-1}$ & $A_n$ \\  \cline{3-4}
	& &$C_{2n}, |C_{2n}|=2n$ &$A_{2n-1}$  &\\  \cline{3-4}
	&  &$BD_{2n}, |BD_{2n}|=4n$ &$D_{n-2}$ &\\ \hline
	(0,0,0) &   &  &  & $D_{2,1,\lambda}$ \\  \hline
					\end{tabular}
					\label{tab:s}
\end{table}
\end{scriptsize}

\section{McKay correspondence} \label{sectM}

McKay correspondence assigns to finite subgroups
of $SU(2)$ group  Dynkin diagrams of some simple Lie algebras  
in the following way. 
Let $G$ be an arbitrary finite subgroup of the group $SU(2)$ and
let $V$ be the restriction of 2-dimensional representation of 
$SU(2)$ on that subgroup $G$. 
Let $\{V_i\}$ be the set of all irreducible 
representations of group $G$, including trivial one. Then
consider decomposition 
\begin{equation}\label{graph1} 
V \otimes V_i = \sum_{j} m_{ij}V_j\,. 
\end{equation}
One can prove that for all pairs $(ij)$, $m_{ij}$ are symmetric, i.e.
$m_{ij}=m_{ji}$, and that coefficients $m_{ij}$ are equal to $0$ or $1$.
Thus one comes to graphs 
$\G_G$ with vertices corresponding to spaces $V_i$ 
and edges between vertices
$V_i,V_j$ iff $m_{ij}\not=0$. 
These graphs appear to be
 Dynkin diagrams of all affine untwisted Kac-Moody algebras of types
$\hat{A}, \hat{D}, \hat{E}$, (McKay, 1980, \cite{McK80}).

Particularly, among finite groups $G$ there are subgroups
of $SU(2)$ which 
are double coverings
(through the double covering $SU(2) \rightarrow SO(3)$)
 of finite subgroups of $SO(3)$ which
are groups of rotational   
symmetries of Platonic solids. These are 

\begin{itemize}

\item 
$2I$ --- binary icosahedral group, which is double covering  
of icosahedral group $I$ - the group of rotational 
symmetries of icosahedron and dodecahedron. It has $2\times 60=120$ elements. 

\item
 $2O$ --- binary octahedral group, double covering  of 
octahedral group O - the group of rotational symmetries 
of cube and octahedron. It has $2\times 24=48$ elements.

\item
$2T$ --- binary tetrahedral group, double covering (extension) of 
tetrahedral group $T$ - the group of rotational symmetries of 
tetrahedron. It has $2\times 12=24$ elements.

\end{itemize}

Besides these three groups, there are only   
two families of finite subgroups of $SU(2)$: $C_{n}$
and $BD_{2n}$. They correspond to 
degenerate Platonic solids, 
namely regular $n$-polygons, which can be considered
as ``polyhedrons'' with two faces, $n$ vertices and $n$  edges.

\begin{itemize}

\item
$C_n$ --- cyclic group, preimage 
 of cyclic group  $C_n$ of rotations of regular n-polygon, if $n$ 
is even, or  $C_n$ --- preimage of cyclic group  $C_n$ of rotations of regular n-polygon, if n is odd, or
 $C_{2n}$ --- preimage of cyclic group  $C_n$ 
of rotations of regular $n$-polygon, if $n$ is odd.
 
\item
$2D_{2n}=BD_{2n}$, binary dihedral group $BD_{2n}$, 
double covering of dihedral 
group $D_{2n}$ of rotations  and reflections  of $n$-polygon.  
(Reflections of $n$-polygon can be considered as rotation in $3$-dimensional
space.)  Group $D_{2n}$ has $n+n=2n$ elements and 
group $BD_{2n}$ has $2\times 2n=4n$ elements.
\end{itemize}

McKay correspondence assigns to all these subgroups
  simple Lie algebras whose Dynkin diagrams
are defined by relation (\ref{graph1}). They are listed in 
third and fourth columns of table \ref{tab:s}.

\section{Diophantine classification of simple Lie algebras.} \label{sectD}

Now turn to the Diophantine classification of simple Lie algebras 
obtained in \cite{M16}. 
The idea is based on the analysis of the so 
called universal character of adjoint representation. 

Recall briefly the main ideas of the universality 
approach to simple Lie algebras \cite{V0,V2}.
Vogel plane is a space which provides coordinatization of
simple Lie algebras. 
It is $2$-dimensional projective space $\C P^2$ factorized
by the action of group  $S_3$ of permutations of  homogeneous coordinates
$(\alpha,\beta,\gamma)$ on $\C P^2$.
  Functions on Vogel plane
 are functions on coordinates
$(\alpha,\beta,\gamma)$, which are scaling invariant and symmetric.  To every simple Lie
algebra corresponds the separate point on Vogel plane.
The values of Vogel parameters (defined up to a rescaling and permutations)
for all simple Lie algebras are given in the table \ref{tab:vogel}. 
All exceptional 
algebras belong to the line  $Exc(n)$ and  are parameterized by numbers
   $n=-2/3,0,1,2,4,8$ 
for exceptional algebras $G_2$, $D_4$, $F_4$,
$E_6$, $E_7$ and $E_8$, respectively.
 
 \begin{scriptsize}
\begin{table}[ht]  
\caption{Simple Lie algebras on Vogel's plane}
	\centering
		\begin{tabular}{|r|r|r|r|}  \hline
   Algebra &$\a$&$\beta$&$\gamma$\cr\hline
     $sl(N)$ &   $-2$  &  $2$    &      $N$      \cr\hline
     $so(N)$ &   $-2$  &  $4$    &     $N-4$     \cr\hline
     $sp(N)$ &   $ -2$ &  $1$    &     $N/2+2$    \cr\hline
    $Exc(n)$ &   $-2$  &  $2n+4$ &     $3n+6$     \cr\hline
	           \end{tabular}
	\label{tab:vogel}
\end{table}
\end{scriptsize}

\noindent   
    
  We say that a function $f$ universalizes some quantity
if  $f$ is an `reasonable' function on Vogel plane which 
is equal to this  quantity at the points of Vogel plane
corresponding to given simple Lie algebras (see table \ref{tab:vogel}). 
For example consider such a quantity as dimension of Lie algebra.
One can see that  function
       $$
d(\alpha,\beta,\gamma)=
   \frac{(\alpha-2t)(\beta-2t)(\gamma-2t)}{\alpha\beta\gamma}\,,
  \quad {\rm where}\quad t=\alpha+\beta+\gamma\,,
       $$
is universal function for dimension of Lie algebra. The value
of this functions at coordinates of an arbitrary
 simple Lie algebra 
(see  table \ref{tab:vogel})
is equal to the dimension of that simple Lie algebra.

    We come to another very important universal function
considering such a quantity as 
a values of character of adjoint representation at
the Weyl line. 
Namely,  let $\gg$  be a simple Lie algebra, and let
$\chi^{(\gg)}_{ad}$ be character of its adjoint representation.
Let $\{\mu\}$ be a set of roots of Lie algebra $\gg$.
Then consider  Weyl vector which is equal to half-sum of all positive roots:
$\rho=\frac{1}{2}\sum_{\mu >0}\mu$. 
Consider  a function 
           $
   f(x)=f_\gg(x)=\chi^{(\gg)}_{ad}(x\rho)
          $\,.
One can see that for this function the following relation holds:
         \begin{equation}\label{weylline}
 f_\gg(x)=\chi^{(\gg)}_{ad}(x\rho)=r+\sum_{\mu}e^{x(\mu,\rho)}\,,       
         \end{equation}
where  $r$ is rank of algebra $\gg$.

  In the papers \cite{MV,Westbury} it was 
suggested the universalisation 
of  function (\ref{weylline}):

\begin{eqnarray} \label{fff}
f(x)=f(x|\alpha,\beta,\gamma)=
\frac{\sinh(x\frac{\alpha-2t}{4})}{\sinh(x\frac{\alpha}{4})}
\frac{\sinh(x\frac{\beta-2t}{4})}{\sinh(x\frac{\beta}{4})}
\frac{\sinh(x\frac{\gamma-2t}{4})}{\sinh(x\frac{\gamma}{4})}\,.
\end{eqnarray}

The values of this function at points of Vogel plane
corresponding to an arbitrary simple Lie algebra $\gg$ 
are equal to character of adjoint representation of the algebra
$\gg$ on Weyl line:
         \begin{equation}\label{diosource}
  \chi^{(\gg)}_{ad}(x\rho)=f(x|\alpha,\beta,\gamma)\,.  
           \end{equation}

  This function is very important for construction of universal expression
for volumes of groups and partition function of Chern-Simon theory
(see \cite{M13}).  
Moreover it turns out that  this function  can be used for Diophantine 
classification of simple Lie algebras, \cite{M16}.
Let's briefly recall results of \cite{M16}.

Left hand side  of equation (\ref{diosource}) 
is the finite sum of exponents,
which happens if all possible poles of  right hand side
of this equation  
are canceled by the zeros of numerator at the points of Vogel plane
corresponding to simple Lie algebra. One can put things upside down 
and seek all points on the Vogel's plane for which this happens. 
  One of the possible patterns of cancellation is the following
  \begin{equation}\label{4p}
    \begin{cases}
      2t-\alpha=(k-1)\alpha \cr
      2t-\beta=(n-1)\beta \cr 
      2t-\gamma=(m-1)\gamma\cr
     \end{cases}\,,\qquad t=\alpha+\beta+\gamma\,,
        \end{equation}
where $k, n, m$ are integers.
 Generally it can be seven such patterns. 
The pattern (\ref{4p}) requires that determinant  
 of corresponding $3\times 3$ matrix vanishes.  
Thus we come to the condition
\begin{eqnarray} \label{abg}
knm=2kn+2km+2nm\,.
\end{eqnarray}
We obtain our main equation (\ref{Dio}) 
in a more general form (\ref{Dio2}).
 If condition (\ref{abg}) is obeyed and all integers are non-zero, then
solutions of equations (\ref{4p})  are
 \begin{eqnarray} 
\label{abg3}
\alpha=\frac{2t}{k}, \, \beta=\frac{2t}{n}, \, \gamma=\frac{2t}{m}\,.  
\end{eqnarray}

Besides that, the non-singular form of Diophantine 
equation (\ref{Dio}), equation (\ref{abg}), has solution $(0,0,m)$, 
with an arbitrary $m$. Solution with $m=0$ is particularly interesting, 
and it is the most general in a certain sense. It corresponds 
to an  object with $E=0$, and hence $V+F=2$. Corresponding solution for Vogel's parameters is  an arbitrary triple $\alpha, \beta, \gamma $ with only restriction $\alpha + \beta + \gamma=0 $. This solution corresponds to superalgebra $D_{2,1,\lambda}$, see  \cite{V0}.

The list of solutions, with corresponding Vogel's parameters and 
simple Lie algebras, is given in the tables \ref{tab:series} and \ref{tab:4abg} below,
series solutions in table \ref{tab:series}, 
and  isolated solutions in table \ref{tab:4abg}. 
Simple Lie (super)algebras, corresponding, in Diophantine classification, to solutions from the first column of table \ref{tab:s},  are presented in the last, fifth column of table \ref{tab:s}.

\begin{scriptsize}
\begin{table}[ht]
\caption{Points in Vogel's plane: series}
	\centering
		\begin{tabular}{|r|r|r|}  \hline
		$\mathfrak{g}$&$\alpha,\beta,\gamma$&k,n,m \\ \hline
		$\mathfrak{sl}(n+1)$ &-2,2,n+1&-n,n,2\\ \hline
	$D_{2,1,\lambda}$&$\alpha+\beta+\gamma=0$&0,0,0\\ \hline
					\end{tabular}
	\label{tab:series}
\end{table}
\end{scriptsize}

\begin{scriptsize}
\begin{table}[ht]
\caption{Isolated solutions }
	\centering
		\begin{tabular}{|r|r|r|r|r|} \hline
		k n m& $\alpha \beta \gamma$ & Dim & Rank & Algebra\\ \hline
5	3	-30	&	-6	-10	1	&	248	&	8	&	$	\mathfrak{e}_8	$\\
4	3	-12	&	-3	-4	1	&	78	&	6	&	$	\mathfrak{e}_6	$\\
3	3	-6	&	-2	-2	1	&	28	&	4	&	$	\mathfrak{so}(8)	$\\
1	-4	-4	&	4	-1	-1	&	0	&	0	&	$	0d_3	$\\
1	-3	-6	&	6	-2	-1	&	0	&	0	&	$	0d_4	$\\
6	6	6	&	1	1	1	&	-125	&	-19	&	$	Y_1	$\\
10	5	5	&	1	2	2	&	-144	&	-14	&	$	Y_{10}	$\\
8	8	4	&	1	1	2	&	-147	&	-17	&	$	Y_{11}	$\\
12	6	4	&	1	2	3	&	-165	&	-13	&	$	Y_{15}	$\\
20	5	4	&	1	4	5	&	-228	&	-10	&	$	Y_{29}	$\\
12	12	3	&	1	1	4	&	-242	&	-18	&	$	Y_{31}	$\\
15	10	3	&	2	3	10	&	-252	&	-8	&	$	Y_{35}	$\\
18	9	3	&	1	2	6	&	-272	&	-14	&	$	Y_{38}	$\\
24	8	3	&	1	3	8	&	-322	&	-12	&	$	Y_{43}	$\\
42	7	3	&	1	6	14	&	-492	&	-10	&	$	Y_{47}	$\\
\hline
	\end{tabular}
	\label{tab:4abg}
\end{table}
\end{scriptsize}

\section{Comparison of McKay correspondence and Diophantine classification.}\label{sectComp}

Now we turn to comparison of connection between solutions and 
simple Lie algebras in McKay correspondence and in Diophantine 
classification. All necessary information is already 
combined in table \ref{tab:s}, so we have to look on its rows. 

We first notice that both approaches combine Platonic solids in dual pairs: 
dodecahedron with icosahedron, cube with octahedron, 
and tetrahedron is alone as it is self-dual. 
Duality is standard, 
including  vertex $ \leftrightarrow $ face correspondence, etc.

The corresponding algebra is the same in both routes, in the first case 
of dodecahedron/icosahedron solution  $(5,3,-30)$, and is the largest 
of exceptional algebras, $E_8$. Two other cases give different outputs
in McKay and Diophantine classification: solution $(4,3,-12)$ 
gives $E_6$ (instead of $E_7$ in McKay correspondence), 
solution $(3,3,-6)$ gives $SO(8)$ (instead of $E_6$ in McKay correspondence). 

Next we see that Diophantine classification choose one of 
two series corresponding to polygon solution $(2,n,-n)$ in 
McKay correspondence, namely $A_{n}$ series (i.e. $sl(n+1$) algebras). 
In McKay correspondence for that degenerate "Platonic solid" 
we get either $A_{n-1}, A_{2n-1}$ or $D_{2n}$. So, there is no 
exact coincidence for polygon solution.

\section{Equivelar maps and solutions 
with positive $k, n, m$.} \label{sectRM}

In equation (\ref{pla1})
we considered  
solutions of Diophantine equation (\ref{Dio}) with  signs
$(++-)$. These solutions correspond to Platonic solids.
There is a number of solutions of equation (\ref{Dio}) 
with `wrong', i.e. all positive, signs. They are 
presented in last 10 lines of table \ref{tab:4abg}, 
and correspond to the so called Y-objects \cite{M16}. E.g. triple $(6,6,6)$
is a solution  of equation (\ref{Dio}), and it
 cannot be interpreted 
in terms of Platonic solids.
However, we can get a reasonable interpretation 
of these solutions 
assigning to them so called equivelar maps
on double torus (compact Riemannian surface of genus $g=2$,
 and Euler characteristics $\chi=-2$).

 {\it Equivelar map} 
on compact Riemann surface  is \cite{McS02} polyhedral map 
with faces all having equal (say $p$) edges and vertices all 
having equal (say $q$) number of edges. 
Initially theory of equivelar maps were developed under 
the name of {\it regular maps} in \cite{E22,T32}. 
Another names for
 equivelar maps are: {\it equivelar $\{p,q\}$ maps}, 
or simply  {\it $\{p,q\}$ maps}, 
where $\{p,q\}$ is  Schl\"{a}fli symbol, 
and {\it locally regular maps} (see below).

In the same way as for equation (\ref{pla1})
one immediately derives for an arbitrary  equiveral map
on double torus an equation
    \begin{equation}\label{pladoubletorus}
     \frac{1}{p}+\frac{1}{q}+\frac{1}{E}=\frac{1}{2}\,.
      \end{equation}
Thus we come to  equation (\ref{Dio}) with positive $k,n,m$. Correspondingly, we get a connection of Y-objects (which have some features of simple Lie algebras, see \cite{M16}) with equivelar maps on genus two surfaces. 

Note that there is a notion 
of {\it regular maps}, 
which are equivelar maps with additional transitivity 
requirement on automorphism group of the graph of the map. 
Actually  transitivity itself already implies that
 the map is equivelar. 
(See \cite{CM80,BS97} for definitions and discussion.) 
Accordingly, another name for equiveral maps is 
{\it locally regular maps}. 
Among solutions of equation (\ref{pladoubletorus}), 
listed in table \ref{tab:4abg}, only $Y_1, Y_{10}, Y_{11}, Y_{15}$ 
and  $Y_{43}$ give rise to regular maps (see  table 9 in \cite{CM80}).

\section{Conclusion}

In this paper we present some observations, which 
connect Diophantine equations (\ref{Dio}), (\ref{Dio2}) 
with Platonic solids, McKay correspondence, equiveral maps 
and Diophantine classification of simple Lie algebras. Latter, in turn, is based on Vogel's universality approach. 
It appears that there are two routes from these equations 
to simple Lie algebras, and they have some similar features 
and some differences. Particularly, maximal exceptional $E_8$ algebra has the same Diophantine solution origin in both routes.  We also observe the connection of  
these equations with equivelar and regular maps on surface of genus $2$
(double tori), and in this way get some interpretation of $Y$-objects. These objects  share some features with simple Lie algebras.

 One can ask on the similar 
interpretation of other six equations \cite{M16} of Diophantine 
classification:

\begin{eqnarray}
kmn&=& mn+2kn+2km \\
kmn&=& mn+2kn+2km+2n-2k\\
kmn&=& mn+kn+2km+3n+2k\\
kmn&=& mn+2kn+2km+2n+2m-3k-5\\
kmn&=& 2mn+2kn+2km-2n-3m\\
kmn&=& 2mn+2kn+2km-2n-2m-2k+5
\end{eqnarray}
Origin of these equations are different patterns of cancellation of zeros in denominator and numerator in universal character (\ref{fff}), so that final answer is regular function in the entire complex $x$ plane, as is the case for all simple Lie algebras. 
One can try to compare them with Artur Caylay's \cite{C1859} form of Euler theorem for some polyhedrons: 

\begin{eqnarray}
d_v V-E+d_f F= 2 D\,,
\end{eqnarray}
where $d_v, d_f, D$ are called vertex figure density, face density and density, respectively. 

Another possibly relevant remark is that for a given number of 
edges, vertices and faces, one can construct 
another polyhedrons, sometimes with the same Euler 
characteristics, so called Kepler-Poinsot 
polyhedrons \cite{C1859,KePo}, which however are not convex. 

Hopefully on the bases of observations in present work one can extend the area of the common Diophantine equations origin of different objects, such as simple Lie algebras, equiveral maps on surfaces of different genus, etc. Another important direction is  an interpretation  (identification) of  $Y$-objects, and their deeper understanding, which is  an interesting challenge.

\section{Acknowledgments.}

We are indebted to MPIM (Bonn), where this work is done, 
for hospitality in autumn - winter 2015-2016. 
HK is grateful to Anna Felikson for encouraging discussions.
Work of RM is partially supported by Volkswagen Foundation 
and by the Science Committee of the Ministry of Science 
and Education of the Republic of Armenia under contract  15T-1C233.

\end{document}